\documentclass[a4,11pt]{article}
\usepackage[intlimits]{amsmath}\usepackage{amssymb}\usepackage{xspace}

\def\lb{\label}
\def\be{\begin{equation}}
\def\ba{\begin{eqnarray}}
\def\ea{\end{eqnarray}}

\def\bb{\bibitem}

\def\e{{\rm e}}
\def\b{{\bar b}}

\begin{document}
\title{
   \begin{flushright} \begin{small}
     LAPTH-902/02  \\ DTP-MSU/02-10 \\
  \end{small} \end{flushright}
\vspace{1cm}
{\bf F0 fluxbranes,  F-walls and new brane worlds} } \vspace{1cm}
\author{
   {\bf G\'erard Cl\'ement$^{a}$}
\thanks{Email: gclement@lapp.in2p3.fr}
 and {\bf Dmitri Gal'tsov$^{a,b}$}
\thanks{Email: galtsov@grg.phys.msu.su} \\ \\
$^{a}$Laboratoire de  Physique Th\'eorique LAPTH (CNRS), \\
B.P.110, F-74941 Annecy-le-Vieux cedex, France\\
$^{b}$Department of Theoretical Physics,\\
     Moscow State University, 119899, Moscow, Russia,}

\date{30 August 2002}
\maketitle
\begin{abstract}
We complete the list of  fluxbrane solutions in classical
supergravities by introducing Melvin type space-times supported by
antisymmetric forms of rank $D-1$ and their pseudoscalar
duals. In IIB theory these solutions belong to the same family as
the seven-brane and D-instanton. In  current notation, 
a fluxbrane supported by the $D-1$ form is an F0 brane, 
its euclidean continuation is a cylindrical background which 
``interacts'' with a pointlike instanton. The general F0 brane can 
have a transverse space $S_k\times R^{(D-k-2)}$ with 
$1\leq k\leq D-2$. For $k=1$ we find the complete solution 
containing four parameters, three of them associated with pointlike 
singularities on the Melvin-type background. The S-dual to 
the F0 brane in ten dimensions is the F8 fluxbrane of 
codimension one, or F-wall, similar fluxbranes exist
in any dimensions if an appropriate form field is present.
F-walls contain naked singularities unless one introduces 
source branes. In such a way one obtains new brane-world type 
solutions with two bulk scalar fields. Their relation to the 
supersymmetric brane-worlds is discussed.

\end{abstract}
\section{Introduction}
Multidimensional generalizations of the Melvin magnetic universe
\cite{Me64}, currently known as fluxbranes, were suggested by Gibbons
and Witshire in 1986 \cite{GiWi86} and generalized for the presence of
dilaton by Gibbons and Maeda \cite{GiMa88}.  Their relevance in
quantum theory was emphasized in \cite{DoGaKaTr94,DoGaGiHo95} where
quantum creation of branes in the fluxbrane background was
considered. More recently fluxbrane solutions were found to be
important in the superstring theory
\cite{Ts95,RuTs01,GaRy98,ChGaSh99b,CoGu00,
Sa01a,GuSt01,CoHeCo01,Em01,Sa01b,Br01,Mo01,Su01,Ur01,ChGaSa01,FiSi01,
TaUe01,Da01,DuMo01,Iv02,Su012,Gu02}. Their applications include
construction of  exact string backgrounds \cite{Ts95,RuTs01},
demonstration of duality between type 0A and IIA string theories
\cite{CoGu00}, brane nucleation \cite{DoGaGiHo95,Gu02}, tachyon
condensation \cite{Da01,Su012,Gu02} and others. Some of  these
solutions (but not all) are obtainable by compactification  of vacuum
spaces on twisted circles \cite{DoGaGiHo95}, other  were found as
exact solutions of supergravity equations. A rather complete list of
(non-intersecting)  fluxbrane solutions in classical ten and
eleven-dimensional supergravities was given in \cite{ChGaSa01}, this
includes F6 and F3 branes in $D=11$ and solutions corresponding to
rank $1-7$ antisymmetric potentials in IIA and IIB theories. The
transverse space of F$p$ fluxbranes may have spherical or cylindrical
topology $S_k\times R^{(8-p-k)}$. Analytic solutions of the Melvin
type exist only for $k=1$, in other cases regular solutions were found
numerically.

Here we add to the list of one-component fluxbranes new solutions
corresponding to so-called exotic branes. In type IIB theory
these are associated with the Ramond-Ramond scalar and its dual
eight-form potential, which give rise to the D-instanton and the
seven-brane \cite{GiGrPe95,GrGu97,EiZa00}. 
Another exotic brane is the
eight-brane, usually interpreted as belonging to massive IIA
supergravity \cite{BrPe99}. An alternative interpretation uses 
a formulation of type II supergravities including a nine-form
potential \cite{BeRo96}, note also the more recent ``brane and
bulk'' formulation of ten-dimensional supergravities including
all Ramond-Ramond potentials of rank $p=0,...,9$
\cite{Be01}. 

Magnetic fluxbranes in IIB theory carrying the flux of an
eight-form, according to current notation \cite{GuSt01}, are 
F0 branes, with a pointlike support, so they present a new 
type of Lorentzian solutions of physical interest. It may seem
counterintuitive to think about such configurations as fluxbranes, 
but in fact they preserve the main fluxbrane property --- the 
finiteness of the form flux, and the space is still of the Melvin type
in the cases where analytic solutions exist. Another natural question 
is that of the object with which the F0 branes are entitled to interact. 
Similarly with F1 fluxbranes which interact with a point charge, 
by dimensional counting F0 fluxbranes should interact with an
instanton, in other words, putting the instanton
on the (euclidean continuation of the) F0 fluxbrane will modify
the D-instanton action. This is, however, not in the
stream of the present investigation, so we do not discuss it further.
For recent generalizations
of instanton solutions to arbitrary dimensions (with the same 
action as ours) and to (A)dS spaces see \cite{GuSa02}.

We show that dilatonic F0 branes exist in any dimension $D$ 
with transverse spaces $S_k\times R^{(D-k-2)}$ for $1\leq k\leq D-2$,
and  we find analytic solutions for $k=1$ and $k=D-2$.
In the first case our solution is generic and contains (after gauge 
fixing) four arbitrary parameters. For certain values of these 
parameters one has a regular space of the Melvin type, 
while genericaly pointlike naked singularities are present. 
In particular, in $D=3$ this is a static Melvin-type
magnetic universe supported by the Maxwell field. In four
dimensions such a solution is supported by the Kalb-Ramond field
which can be equally associated with the NS potential.

Electric fluxbranes associated with an eight-form in  ten
dimensions are F8 branes,  alternatively they can be viewed as
magnetic solutions supported by a RR scalar. These 
have the well-known structure of domain walls.
F-walls turn out to be singular on some nine-dimensional
hyperplane, so it is suggestive to introduce the source
eight-brane(s). We show that a single brane of negative
tension matches indeed to the F-wall producing an otherwise
non-singular space-time. The situation is similar in any
dimension in the presence of a scalar potential. 
It is worth noting that
the source brane does not carry the charge of the fluxbrane form
field, which in this case is not of the correct rank. 
To compare the situation
with the recently studied charged supersymmetric branes
\cite{AlMeOr00,BeKaPr00,Li00,GrQuZaTa01}, we also construct 
a solution for a source
eight-brane carrying the RR charge of a nine-form potential.
This is similar to the charged Randall-Sundrum flat brane-world
without bulk cosmological constant, its five-dimensional
counterpart is reminiscent of the solutions discussed in
\cite{KaKoLiTs01}. Our solution containing a single source brane is
regular outside the brane in the case of a negative brane tension.
More complicated regular solutions should exist, including 
several source branes
of both positive and negative tensions.

\setcounter{equation}{0}
\section{ F0 branes}
In this paper we will essentially use the formulation of Ref.
 \cite{ChGaSa01}. Consider the coupled system of
$D$-dimensional gravity, an antisymmetric form of rank $D-1$
and the dilaton: \be\label{Sq}
 S = \int \left( R - \frac12\,
\partial_\mu \phi\,
\partial^\mu \phi - \frac{1}{2\cdot (D-1)!}  \, {\rm e}^{a\phi} \,F_{[D-1]}^2
\right)\sqrt{-g}\,d^{D} x, \end{equation} 
where $F_{[D-1]} = dA_{[D-2]}$ and
$a$ is an arbitrary dilaton coupling.
The F0-brane spacetime we are looking for can be parametrized by
three functions $A(r),\, B(r),\, C(r)$ of the radial variable as
follows \be\label{ds} ds^2 = -{\rm e}^{2A} dt^2   +  {\rm e}^{2B}
dr^2 + {\rm e}^{2C} (r^2 d\Omega_k^2 + \delta^{mn}\,dy_m
\,dy_n),\end{equation} with $\Omega_k,\, (1 \le k \le D-2)$, being the
spherical volume element, and $m, n=1,..., D-2-k$. The  F0-brane
is centered at the origin, the ($D-1$)-dimensional transverse
space is the product of the $k+1$-dimensional ball and  the flat
euclidean space of dimension $D-2-k$. The magnetic ansatz for the
F0-brane reads \be\lb{ansF} F_{[D-1]} = f(r)\, dr \wedge
\epsilon_{[k]} \wedge dy_1 \wedge \cdots \wedge dy_{D-2-k},\end{equation}
where $\epsilon_{[k]}$ is the  volume element of the unit
$k$-sphere. The $D-1$-form field equation   \be\label{dF}
\partial_\mu \left( \sqrt{-g} \, {\rm e}^{a\phi} \,
F^{\mu\nu_1\cdots\nu_{9}} \right) = 0,\end{equation} is then easily solved
with the above ansatz to give \be\lb{f}f = 2 \b  \, {\rm e}^{ H +
2(B - A) -a \phi},\end{equation} where $\b$ is an integration constant, and
\be\lb{H}H = A - B + (D-2)C + k \ln r.\end{equation}

The non-zero components of the Ricci tensor for the metric
(\ref{ds}) are given in \cite{ChGaSa01}, the resulting Einstein
equations read \ba
A''+ H'A' - 2 \b^2  {\rm e}^{2(B - A) -a \phi}&=&0,\lb{EqA}\\
C'' + H'C' + r^{-2}\left(r H' - 1 - (k-1) {\rm
e}^{2B-2C}\right)&=&0,
\lb{EqH}\\
(A'' + {A'}^2 - A'B') +  (D-2) (C'' + {C'}^2 - B'C') &&\nonumber\\ + k
r^{-1}(2C'-B') + \frac12 {\phi'}^2  &=& 0. \lb{EqR} \ea In addition,
for all $k < D-2$ the vanishing of the $R_{yy}$ component of the
Ricci tensor leads to the equation \be\lb{C} C'' +  H'C'   = 0.
\end{equation}
The dilaton equation is similar to (\ref{EqA}):\be\lb{Eqphi}
\phi'' + H'\phi' = 2 a \b^2\,{\rm e}^{2(B - A) -a \phi }.\end{equation}

\subsection{$k=1$ }

We start with the simplest case $k=1$ (assuming $D>3$), in which
the system of equations can be integrated in a closed form. In
this case, Eq. (\ref{EqH}) combined with (\ref{C}) gives \be
\lb{H1} H = \ln(r/l)  ,\quad  l=\rm const. \end{equation} Substituting this
into Eq. (\ref{C}) we find \be\lb{solC} C =  \gamma \ln
({r}/{r_0}), \quad  r_0=\rm const , \end{equation} and consequently
\be\lb{solB} B=A+\gamma(D-2) \ln  ({r}/{r_0}) + \ln l. \end{equation}
Integrating a linear combination of Eqs.
(\ref{EqA}) and (\ref{Eqphi}) we find the following relation between
$\phi$ and $A$: \be\lb{Aphiq}\phi = a A + \beta \ln(r/r_1),\quad
\beta, r_1 = \rm const.\end{equation} Using this together with the two
previous relations in the Einstein equation (\ref{EqR}), we
arrive at the following non-linear equation for the dilaton
function \be\lb{pheqq} \ddot\phi + \frac{a}2\dot\phi^2 -
2\alpha\dot\phi
 + 2\alpha\beta - a\frac{D-3}{D-2}(\alpha^2-1)  = 0,
\end{equation} where the dot denotes the derivative with respect to the
logarithmic variable $\tau=\ln r$, and $\alpha = 1 + \gamma(D-2)$.
This is a first order equation for $\dot\phi$, which is solved by
\be\lb{phisolq}
\phi=\phi_0 + \frac2{a}\left[\alpha(\tau+\tau_0) +
\ln\left(2\cosh \kappa(\tau + \tau_0)\right)\right],\end{equation} with
\be\lb{kappa} \kappa^2 = \alpha^2 - a\beta\alpha +
\frac{a^2}2\frac{D-3}{D-2}(\alpha^2-1).\end{equation} Here $\phi_0$ and
$\tau_0$ are two new integration constants, and it is convenient
to rename \be \tau_0=\ln b. \end{equation}

So far we have not used the Einstein equation (\ref{EqA}).
Substituting (\ref{Aphiq}), (\ref{phisolq}), (\ref{solC}) and (\ref{solB})
into (\ref{EqA}), one finds \be \b=\frac{2\kappa
b}{|a|l}(br_0)^{\alpha-1}\e^{a\phi_0/2}. \end{equation} As a result, we
obtain the following solution \ba {\rm e}^{2A} &=& {\rm
e}^{2\phi_0/a}\left(br\right)^{4(\alpha-\kappa)/a^2}
\left(\frac{r_1}{r}\right)^{2\beta/a}\left(1 +
\left(br\right)^{ 2\kappa}\right)^{4/a^2},\\
{\rm e}^{2B} & = & l^2\left(\frac{r}{r_0}\right)^{2(\alpha-1)}
{\rm e}^{2A},\\{\rm e}^{2C} & = &
\left(\frac{r}{r_0}\right)^{2(\alpha-1)/(D-2)},\\ {\rm e}^{a\phi/2
} &=& {\rm e}^{a\phi_0/2}\left(br\right)^{\alpha -\kappa}
\left(1 + \left(br\right)^{ 2\kappa}\right),\\
f &=& \frac{4 \kappa}{|a|} {\rm e}^{-a\phi_0/2}(br_0)^{1-\alpha}
 \left(br\right)^{2\kappa - 1}
\left(1 + \left(br\right)^{ 2\kappa}\right)^{-2},
 \ea
depending on arbitrary constants $r_0, r_1, l, b, \alpha,\kappa,
\phi_0$ ($\beta$ being related to $\kappa$ by (\ref{kappa})).
Without loss of generality, one can fix the scale of time and of
the radial coordinate $r$, and choose a normalization point for
the dilaton so that \be r_0 = r_1 = b^{-1}, \quad \phi_0 = 0, \end{equation}
so finally we have a four-parametric family of solutions.
Evidently, $b$ is the fluxbrane field strength parameter, the
physical sense of other parameters is yet to be clarified.

Remarkably, the magnetic field $F$ depends only on the parameters
$b$ and $\kappa$. It derives from the potential \be A_{[D-2]} =
\frac2{|a|b}\,\frac{(br)^{2\kappa}}{1+(br)^{2\kappa}}\, d\varphi
\wedge dy_1 \wedge \cdots \wedge dy_{D-3}\,. \end{equation} The
corresponding magnetic flux per unit (D-3)-dimensional volume is
finite, \be \Phi = \oint_{r \to \infty}A_{[D-2]}  =
\frac{4\pi}{|a|b} dy_1 \wedge \cdots \wedge dy_{D-3}\,. \end{equation}

Solutions regular in the origin are obtained for $\gamma = 0$
($\alpha = 1$), $\beta = 0$ ($\kappa=1$) and $l = 1$. The
resulting metric \be\label{dsM} ds^2 = \left(1 +
b^2r^2\right)^{4/a^2}(- dt^2   +  dr^2 ) +
 r^2 d \varphi^2 + \delta^{mn}\,dy_m \,dy_n\end{equation}
is nothing else than the smeared $(1+2)$-dimensional dilatonic
Melvin-type solution. The corresponding dilaton and the form
field are \ba {\rm e}^{a\phi/2 }
&=&   1 + \left(br\right)^{ 2},\\
F_{D-1} &=& \frac{4 b\kappa r}{|a|} \, \left(1 +
\left(br\right)^{ 2}\right)^{-2} dr \wedge \epsilon_{[k]}\wedge
dy_1 \wedge \cdots \wedge dy_{D-2-k}.
\ea

\subsection{$2\leq k\leq D-3$} 

For $k\neq 1$ there is no hope to find the general solution
analytically (except in the ultimate 
spherically symmetric case $k = D - 2$,
see below), so one is led to
numerical calculations. Still, a particular solution of the type
discussed in \cite{ChGaSa01} can be found by making the ansatz
$C\equiv 0,\; \phi=a A$ which is consistent with the set of
equations. Then one can derive a decoupled second order
differential equation for $A$: \ba\lb{AAq}
\left(2krA''+2k^2A'+4kr{A'}^2-a^2r^2{A'}^3\right)(k-1){\rm
e}^{(a^2+2)A}  &&\nonumber\\+ 2\b^2 r \left(2r^2A''-2r
A'(2k-1)-2k(k-1)+a^2r^2{A'}^2\right) =0&&,\ea while $B$ is given in
terms of $A$ as follows \be  {\rm
e}^{2B}=\frac{2krA'+k(k-1)-a^2r^2{A'}^2/2}{k(k-1)+2b^2 r^2 {\rm
e}^{-(a^2+2)A}}.\end{equation} This system admits an analytic solution
 \ba 2A &=& \gamma \ln
\left(\frac{2 b^2 r^2}{\gamma(\gamma +k-1)}\right), \lb{attract1}\\
 2B&=&\ln\left(1+\frac{\gamma}{k-1}\right),\quad
 \gamma=\frac{2}{a^2+2}.\lb{attract2}\ea
This solution is singular at the origin, but it serves as an
attractor for regular ones that can be found numerically. The
latter are quite similar to more general solutions found in
\cite{ChGaSa01} so we do  not give them here.

\subsection{$k=D-2$ }

Another case in which some progress can be made is that of 
spherical symmetry, $k=D-2$. Putting $C = E - \ln r$, so that
\be\lb{Hsphe} H = A - B + kE, \end{equation} we find that Eqs. (\ref{EqH})
and (\ref{EqR}) take the simpler form \ba && E'' + H'E'  =  (k-1)
e^{2(B-E)}, \lb{EqE} \\&& A'' + {A'}^2 - A'B'+ k(E'' + {E'}^2 - E'B')
+ (1/2){\phi'}^2  =  0 \lb{EqR1}. \ea We see that Eq. (\ref{EqE})
is of the same form as Eqs. (\ref{EqA}) and (\ref{Eqphi})
provided \be\lb{EAphi} E = A + \frac{a}2\phi + \epsilon \end{equation} with
constant $\epsilon$. Furthermore, in the case $k = D-2$ one can
always use radial coordinate transformations to choose a gauge
such that \be\lb{gaugeH} H=\ln(r/l). \end{equation} Note that this condition
does not fix the gauge completely, the condition (\ref{gaugeH})
being, up to an additive constant, invariant under
reparametrizations $r \to \bar{r} = r^p$, which lead to $\bar{H}
= H + (p-1)\ln r + {\rm const} = \ln\bar{r} + {\rm const}$. As in
the case $k = 1$, the comparison of Eqs. (\ref{EqA}) and
(\ref{Eqphi}) now leads to \be \phi = aA + \beta\ln(r/r_1), \end{equation} where
$\beta$ is some constant. On account of (\ref{EAphi}) it follows
that \be E = \gamma^{-1}A + \frac{a\beta}2\ln(r/r_1) + \epsilon, \end{equation} with
$\gamma^{-1} = 1 + a^2/2$. Eqs. (\ref{EqA}) and (\ref{EqE}) are then equivalent
provided
\be
\e^{2\epsilon} = \frac{(k-1)\gamma}{2\bar{b}^2}.
\end{equation}
Inserting these relations in Eq.
(\ref{EqR1}), we arrive at the first order differential equation
with respect to the variable $\tau = \ln(br)$: \be \ddot{A} +
(1-k)a\beta\dot{A} + \frac{\beta^2}2\left(1-\frac{k^2}{k+\gamma}
\right) + \frac{1-k-\gamma}{\gamma}\dot{A}^2 = 0. \end{equation} For $\beta
\neq 0$ (the solution in the case $\beta = 0$ is a gauge
transform of the power-law solution (\ref{attract1}),
(\ref{attract2})), this is solved by \be\lb{solcot} \dot{A} =
\alpha + \kappa\delta\cot\kappa(\tau+\tau_0), \end{equation} with \be \alpha =
(k-1)\delta\frac{a\beta}2, \quad \delta = -
\frac{\gamma}{k+\gamma-1}, \quad \kappa^2 =
\frac{\beta^2}2\left(k-2+\frac1{k+\gamma}\right). \end{equation}

At this point, we may further fix the gauge by choosing $\beta =
-2/a\delta$ so  that $\alpha =1 -k$. Integrating (\ref{solcot})
(with convenient choices for the integration constants) we obtain 
the solution as 
\ba \e^{2A} & = &
(br)^{-2(k-1)}|\sin\kappa\tau|^{-\frac{2\gamma}{k+\gamma-1}},
\lb{Asin}\\
\e^{2B} & = &  |\sin\kappa\tau|^{\frac{-2(k+\gamma)}{k+\gamma-1}},\lb{Bsin}\\
\e^{2C} & = &  b^2\e^{2\epsilon}|\sin\kappa\tau|^{-\frac{2}{k+\gamma-1}}
,\lb{Csin}\\
\e^{a\phi} & = &
(br)^{2k}|\sin\kappa\tau|^{\frac{2(\gamma-1)}{k+\gamma-1}} \lb{phisin}
\ea 
($\tau = \ln(br)$). Inserting these in (\ref{EqA}), we check that this
equation is satisfied provided
\be
{\bar b}^2 =\frac{\gamma\kappa^2}{2(k+\gamma-1)}.
\end{equation} 
The corresponding
magnetic potential is 
\be 
A_{[k]} = -\frac{2\bar b}{\kappa b}\,\e^{k\epsilon}\,
\cot\kappa\tau\, \epsilon_k. \end{equation}

The solution (\ref{Asin})-(\ref{phisin}) is defined only in
disjoint sectors $n\pi < \kappa\tau < (n+1)\pi$. In such a
sector, the spacetime metric behaves near the formal singularity
$\kappa\tau = n\pi$ as 
\be\lb{asing} 
ds^2 \sim -R^{2\gamma}dt^2 + dR^2 + b^2\e^{2\epsilon} R^2
d\Omega_k^2, \quad (R \to \infty) 
\end{equation} 
(we have put $\kappa\tau =
n\pi + R^{-(k+\gamma-1)}$, and neglected irrelevant
multiplicative constants). We see that the behavior near the
singularity is governed by the attractor
(\ref{attract1}), (\ref{attract2}). A first integral
for geodesic motion in such a metric is \be \dot{R}^2 +
\frac{L^2}{R^2} = E^2 R^{-2\gamma} - \sigma \end{equation} ($\sigma = +1$,
0, or $-1$ for timelike, null, or spacelike geodesics). Remembering
that $\gamma = (1+a^2/2)^{-1}
> 0$, we see that timelike geodesics and nonradial ($L \neq 0$) null
geodesics are reflected by a potential barrier before reaching
infinity, while spacelike and radial null geodesics reach
infinity for an infinite value of the affine parameter. The
behavior being similar at the other end $\kappa\tau \to
(n+1)\pi$, it follows that the spacetime
(\ref{Asin})-(\ref{Csin}) is geodesically complete. 

\section{Particular cases}
\setcounter{equation}{0}

\subsection{1+2 dilaton Melvin}
In $1+2$ dimensions the action (\ref{Sq}) corresponds to the
Einstein-Maxwell dilaton theory. In this case, $k=1=D-2$, so the
analytic solution with $k=1$ is the general one. One can also fix
$\alpha=1$ ($C = 0$) as a gauge condition. Then the solution can be
rewritten as follows  \be \lb{ds3}
ds^2=\left(\frac{r}{r_1}\right)^{2(\kappa^2+1)/a^2}
\left[\bigg(\frac{r}{r_b}\bigg)^\kappa+
\bigg(\frac{r_b}{r}\bigg)^\kappa\right]^{4/a^2} \left(-dt^2+l^2
dr^2\right) +r^2d\varphi^2, \end{equation} where $r_b=1/b$ and we removed
the constant dilaton factor by time rescaling. This spacetime is
supported by the following Maxwell two-form \be F=
 \frac{4\kappa}{|a|} {\rm e}^{-a\phi_0/2}\,\frac{r_b}{r}\,
\left[\left(\frac{r}{r_b}\right)^\kappa+
\bigg(\frac{r_b}{r}\bigg)^\kappa\right]^{-2} \;dr\wedge
d\varphi,\end{equation} and dilaton \be {\rm e}^{a\phi/2 } =  {\rm
e}^{a\phi_0/2}\, \frac{r}{r_b}
\left[\left(\frac{r}{r_b}\right)^\kappa+
\bigg(\frac{r_b}{r}\bigg)^\kappa\right]. \end{equation} We note some
overlap in this point with the recent paper \cite{Fe02} which
appeared after our calculations had been completed.

For $\kappa = 0$ we recover the singular solution for a massless
gravitating scalar field in 2+1 dimensions \cite{BBL}, \cite{sig}
\be\lb{k0}
ds^2 = r^{2/a^2}(-dt^2+dr^2) + r^2\,d\varphi^2, \quad \phi = \frac2a
\ln r, \quad F = 0.
\end{equation}
In this case $a$ is an arbitrary integration constant. Up to a
coordinate transformation $r \to r^p$, the behavior
(\ref{k0}) dominates the asymptotic behavior of (\ref{ds3}) ($p = 1+\kappa$)
 \be
ds^2 \sim r^{2(1+\kappa)^2/a^2}(-dt^2+dr^2) + r^2\,d\varphi^2, \quad
\e^{a\phi/2} \sim r^{1+\kappa}, \quad F \sim r\,r^{-2(1+\kappa)}\,dr\wedge
d\varphi,
\end{equation}
as well as the small $r$ behavior ($p=1-\kappa$)
\be
ds^2 \sim r^{2(1-\kappa)^2/a^2}(-dt^2+dr^2) + r^2\,d\varphi^2, \quad
\e^{a\phi/2} \sim r^{1-\kappa}, \quad F \sim r\,r^{-2(1-\kappa)}\,dr\wedge
d\varphi.
\end{equation}
So the general solution (\ref{ds3}) may be viewed as a soliton
interpolating between two ``vacua'' where the magnetic field is negligible.

For $\kappa=1$ we get the regular dilaton-Melvin solution. Coming back
to the initial parametrization, and taking $\phi_0=0$, we obtain 
\ba
\lb{Mel3} ds^2 &=& \left(1+b^2r^2\right)^{4/a^2}
\left(-dt^2+l^2 dr^2\right) +r^2d\varphi^2,\\
 F &=& \frac{4br}{|a|\left(1+b^2r^2\right)^2}
 \;dr\wedge d\varphi,\\
{\rm e}^{a\phi/2 } &=&   \left(1+b^2r^2\right). \ea The flux of
the magnetic field \be \Phi=\int F=\oint_{r\to\infty} A_\varphi
d\varphi, \end{equation} where \be A_\varphi =\frac{2br}{1+b^2r^2}\end{equation} is the
one-form potential generating the two-form field $F=dA$, is finite
\be \Phi= \frac{4\pi}{ab}. \end{equation}

For $a=2$ this solution can be obtained by a twisted
compactification of flat four-dimensional space, similarly to the
case of
usual dilatonic Melvin solutions in higher dimensions. Starting
with \be ds^2_4=-dt^2+dr^2+r^2d\varphi^2+dy^2 \end{equation} and performing
dimensional reduction with respect to the Killing vector \be
K=\partial_y+b\partial_\varphi, \end{equation} one finds \be
ds_4^2=\e^{-\phi}ds_3^2+\e^\phi(dy+A_\varphi d\varphi)^2 \end{equation} where
\be A_\varphi=\frac{br^2}{1+b^2r^2},\quad \e^\phi =1+b^2r^2. \end{equation}
Similarly, the general solution (\ref{ds3}) with $a = 2$ is a twisted
compactification of the cylindrically symmetric Levi-Civita metric \cite{LC}
\be
ds_4^2 = r^{(\kappa^2-1)/2}(-dt^2 + dr^2) + r^{1+\kappa}\,d\varphi^2 +
r^{1-\kappa}\,dy^2. 
\end{equation} 

\subsection{F0 branes in IIB theory}
In the case of IIB theory, $D=10, \,a=-2$ and F0 fluxbranes correspond to the
eight-form potential $A_{8}$ dual to the RR scalar. There are
eight different solutions generated by the form field
\be\lb{ansF9} F_{[9]} = f(r)\, dr \wedge \epsilon_{[k]} \wedge
dy_1 \wedge \cdots \wedge dy_{8-k},\end{equation} with $k=1,\ldots,8$.
We find the following metric for $k=1$: \be
ds^2=(br)^{\alpha+\beta-\kappa}\left(1 + \left(br\right)^{
2\kappa}\right)\left(-dt^2+ l^2(br)^{2(\alpha-1)}\, dr^2\right)+
(br)^{(\alpha-1)/4}\left( r^2 d \varphi^2 + \delta^{mn}\,dy_m
\,dy_n\right),\end{equation} with
$\kappa^2=\alpha^2+2\beta\alpha+7(\alpha^2-1)/4$, while the
dilaton and the form field are \ba   {\rm e}^{-\phi } &=&
\left(br\right)^{\alpha -\kappa}
\left(1 + \left(br\right)^{ 2\kappa}\right),\\
f &=& 2\kappa\frac{(br)^{2( \kappa - 1)}}{ \left(1 +
\left(br\right)^{ 2\kappa}\right)^{2}}.\ea The
$\kappa=\alpha=l=1$ solution regular at the origin reads \ba {\rm
e}^{2A} &=& 1 + b^2r^2,\\{\rm e}^{-\phi } &=&
 1 + b^2r^2 ,\\
f &=& 2b r \left(1 + b^2r^2\right)^{-2}. \ea The corresponding
space-time actually is nothing else than the 1+2 dilaton Melvin metric
smeared in seven extra dimensions \be\label{ds10} ds^2 = \left(1 +
b^2r^2\right)(- dt^2   +   dr^2 ) +
 r^2 d \varphi^2 + \delta^{mn}\,dy_m \,dy_n\,.\end{equation}
Numerical solutions for $k\neq 1$ approach the following
asymptotics:
 \be A = \frac16 \ln \frac{18 b^2 r^2}
{3k-2},\quad B=\frac12\ln\left(1+\frac{1}{3(k-1)}\right).\end{equation}

\subsection{$D=4$ dilaton-axion $F0$ branes}
In four spacetime dimensions, the relevant theory is the
dilaton-axion system coupled to gravity which for $a=1$ is
associated with the toroidally compactified  heterotic string.
Here the metric for the $k=1$ solution has cylindrical symmetry and
is just a smeared $1+2$ solution, while for $k=2$ one has the
spherically symmetric spacetime \be\label{ds4} ds^2 = -{\rm
e}^{2A} dt^2   +  {\rm e}^{2B} dr^2 + r^2 d\Omega_2^2, \end{equation} where
the metric functions $A,\, B$ can be found numerically as
described above. Special solutions are the geodesically complete solutions
\be
ds^2 = -(br)^{-2}(\sin\kappa\tau)^{-\frac{2\gamma}{1+\gamma}}\,dt^2 + 
(\sin\kappa\tau)^{-\frac{2(2+\gamma)}{1+\gamma}}\,dr^2 +
\e^{2\epsilon}\,(br)^2(\sin\kappa\tau)^{-\frac{2}{1+\gamma}}\,d\Omega^2\,,
\end{equation}
with $\tau = \ln r$, $\gamma = (1+a^2/2)^{-1}$, $\kappa =
(1+\gamma)/a\gamma\sqrt{1+\gamma/2}$, $e^{2\epsilon} =
a^2\gamma^2(1+\gamma/2)/(1+\gamma)$, and the singular attractor  
\be
ds^2 = -r^{2\gamma}\,dt^2 + (1+\gamma)\,dr^2 + r^2\,d\Omega^2.
\end{equation}
For $a = 0$ ($\gamma = 1$)  this last
metric has previously been considered by Hayward \cite{Ha00}, in
the context of gravitational collapse of a scalar field (actually
the axion field dual to the three-form); it is continuously
self-similar, and the singularity $r = 0$ is null.
\setcounter{equation}{0}
\section{F-walls}
The S-dual solutions to the F0 branes are F-walls of codimension
one. The most natural way to construct them is to pass to the
dual action \be\label{S1} S_1 = \int \left( R - \frac12\,
\partial_\mu \phi\,
\partial^\mu \phi - \frac{1}{2}  \, {\rm e}^{a\phi} \,F_{[1]}^2
\right)\sqrt{-g}\,d^{D} x, \end{equation} where $F_{[1]} = dA_{[0]}$, with
$A_{[0]}$ being the RR scalar in IIB theory (with $a=2$, for
generality we continue to assume the dimension $D$ and dilaton
coupling $a$ to be arbitrary). Assuming $SO(D-2,1)$ Lorentz symmetry on the
world-volume, we choose the metric in the gauge \be\label{dsw}
ds^2 = {\rm e}^{2A} \eta_{ab} dx^a dx^b   +  dy^2, \end{equation} where $A$
is the function of the transverse coordinate $y$ and $a,\,b
=0,\cdots,D-2$. The form field is simply \be F_{[1]}=f(y)\, dy,
\end{equation} and the solution for $f$ is readily obtained \be\lb{fw} f=2b
{\rm e}^{-(D-1)A - a\phi}. \end{equation} The Ricci tensor has the following
non-zero components: \ba
R_{ab} &=& - \eta_{ab}\left[A'' + (D-1){A'}^2\right]{\rm e}^{2A},\\
R_{yy} &=& - (D-1)(A'' + {A'}^2), \ea and the corresponding Einstein
equations read \ba R_{ab} &=& 0,\\ R_{yy} &=& \frac12\left({\phi'}^2
+ {\rm e}^{a\phi}f^2\right).\ea Therefore, one has a decoupled
equation for the metric function \be\lb{Aeq} A'' + (D-1) {A'}^2 =0,
\end{equation} and a constraint \be\lb{cons} 2(D-1)(D-2){A'}^2 =   {\phi'}^2 +
{\rm e}^{a\phi}f^2. \end{equation} The dilaton equation reads \be
\lb{dilw}\phi'' + (D-1)A' \phi' =\frac{a}2  {\rm e}^{a\phi} f^2.
\end{equation}

It is convenient to substitute \be \lb{AZ} A = \frac1{D-1}\ln Z.
\end{equation} Integration of Eq. (\ref{Aeq}), which is equivalent to $Z''=0$,
gives \be Z= \alpha(y +  y_0), \end{equation} where $\alpha$ and $y_0$ are
two integration constants. Introducing a new independent variable
$\tau$ such that \be\lb{tz} \frac{d\tau}{dy} = \frac{1}{Z}, \end{equation}
the dilaton equation becomes \be\lb{phet} \ddot\phi-2ab^2{\rm
e}^{-a\phi} = 0,\end{equation} while the constraint (\ref{cons}) reads \be
\lb{dilint} \frac{{\dot\phi}^2}{2}+2b^2{\rm
e}^{-a\phi}=\alpha^2\frac{D-2}{D-1}. \end{equation} The solution to Eqs.
(\ref{phet}) and (\ref{dilint}) is \be \phi =
\frac2{a}\ln\left(\frac{ab}{\nu\alpha}\, \cosh\left[\nu\alpha
(\tau+\tau_1)\right]\right),\quad \nu=a\,\sqrt{\frac{D-2}{2(D-1)}},
\end{equation} where $\tau_1$ is an integration constant. Finally,
from (\ref{tz}) the variable $\tau$ is related to $y$ by \be
\tau+\tau_1 = \alpha^{-1}\ln [c(y+y_0)], \end{equation} with a new parameter
$c$, so that the solution takes the form \ba
{\rm e}^{2A} &=& (\alpha(y+y_0))^{2/(D-1)},\\
{\rm e}^{a\phi/2} &=& \frac{ab}{2\nu\alpha}
\left[\left(c(y+y_0)\right)^{\nu}
+\left(c(y+y_0)\right)^{-\nu}\right],\\
f &=&
\frac{8\nu^2\alpha}{a^2b(y+y_0)}\left[\left(c(y+y_0)\right)^{\nu}
+\left(c(y+y_0)\right)^{-\nu}\right]^{-2}.\ea It has a naked
singularity at $y=-y_0$.

The F8 fluxbrane of IIB theory corresponds to $D=10,\, a=2$,
so $\nu=4/3$. Letting for simplicity $\alpha = 1, y_0=0$, we
obtain: \ba
ds^2 &=& y^{2/9} \eta_{ab}dx^a dx^b +dy^2,\\
{\rm e}^{\phi} &=& \frac{3b}{4}
 \left((cy)^{4/3}+ (cy)^{-4/3}\right),\\
f &=& \frac{32}{9by}
 \left((cy)^{4/3}+ (cy)^{-4/3}\right)^{-2}.
\ea After the coordinate transformation  $z=3/4 y^{4/3}$ and a
suitable rescaling of $x^a$, one can cast this solution into the
`Melvin-like' form: \ba
ds^2 &=& z^{1/6} \eta_{ab}dx^a dx^b + z^{-1/2}dz^2,\\
{\rm e}^{\phi} &=& \frac{3b}{4{\bar c} z}
 \left(1+ {\bar c}^2 z^2\right),\\
f &=& \frac{8{\bar c}^2 z^{5/4}}{3b}
 \left(1+ {\bar c}^2 z^2\right)^{-2}.
\ea


\setcounter{equation}{0}
\section{Adding source brane }
The nature of the singularity in the F-wall solutions suggests
that one can try to improve these solutions by introducing a source
brane. In fact, a non-singular metric can be obtained by
orbifolding the above solution at some point away from the
singularity. In physical terms, we add to the lagrangian
(\ref{S1}) a source brane term and (for more generality and to
make contact with the Randall-Sundrum setup) a (negative)
cosmological constant \be S=S_1 - \int\left(2\Lambda +2\lambda
{\rm e}^{\beta\phi} \delta(y)\right)\sqrt{-g}\;d^Dx.\end{equation}  Then,
Eqs. (\ref{Aeq}) and (\ref{dilw}) are replaced by \ba A'' + (D-1)
{A'}^2 &=& -\frac {\lambda_0}{D-2}\,\delta(y)-\frac{2\Lambda}{D-2},\\
\phi'' + (D-1)A' \phi' &=&\frac{a}2 \; {\rm e}^{a\phi}
f^2+2\lambda_0\, \beta\, \delta(y),\ea where
\be\lambda_0=\lambda\;{\rm e}^{\beta\phi(0)},\end{equation} while the
constraint equation now reads \be\lb{cons1} 2(D-1)(D-2){A'}^2
+4\Lambda =   {\phi'}^2 + {\rm e}^{a\phi}f^2. \end{equation} Note that the
form field  does not interact with the brane, so the solution
(\ref{fw}) remains true.

Using again the parametrization  (\ref{AZ}), we find for $Z$ the
equation \be\lb{ZR} Z''-\kappa^2 Z = -
\lambda_0\,\frac{D-1}{D-2}\,Z\;\delta(y),\end{equation} where\be\lb{kala}
\kappa^2=-\frac{2(D-1)\Lambda}{D-2},\end{equation} and $\Lambda$ is assumed
to be negative. For $\beta = 0$ and in the absence of the form
field ($b=0$) it is consistent to set the dilaton to zero. In
this case a particular solution will be the flat RS brane-world
in the AdS  bulk, it is given by \be Z=\alpha {\rm
e}^{-\kappa|y|},\quad \alpha>0,\;\;\; \kappa>0,\end{equation} the matching
condition at $y=0$ leading to \be 2\kappa
=\lambda_0\,\frac{D-1}{D-2}.\end{equation} The tension $\lambda$ of the
brane thus has to be positive.

Adding $F_{[1]}$ and the dilaton (i.e. two bulk  scalar fields)
changes the situation as follows. Introducing  the variable
$\tau$ as in (\ref{tz}), one obtain the dilaton equation \be
\ddot\phi-2ab^2{\rm e}^{-a\phi} = 2\lambda_0\beta\,\delta(y). \end{equation}
The first integral (which does not contain discontinuities)
replacing (\ref{dilint}) is \be\lb{dilintb}
\frac{{\dot\phi}^2}{2} +2b^2{\rm e}^{-a\phi} = 2\mu^2, \end{equation} where
from the constraint equation (\ref{cons1}) \be
\mu^2=\frac{D-2}{2(D-1)}\left({Z'}^2-\kappa^2 Z^2\right), \end{equation}
$\kappa$ being still given by (\ref{kala}). In view of this
relation, the exponential solution of the $Z$-equation  is no
longer valid, instead we should take the following solution to
Eq. (\ref{ZR}):\be Z=\alpha\sinh \kappa(|y|+y_0),\end{equation} so that
\be\mu^2=\frac{D-2}{2(D-1)}\kappa^2\alpha^2.\end{equation} Now the matching
condition on the brane gives \be -2\kappa\coth \kappa
y_0=\lambda_0\,\frac{D-1}{D-2},\end{equation} leading to a negative 
brane tension. The solution to the dilaton equation now reads \be
\phi=\frac2{a}\ln\left[\frac{b}{\mu}\cosh a\mu
(|\tau|+\tau_0)\right], \end{equation} and the matching condition reads \be
\frac{2\mu}{\alpha}\frac{\tanh(a\mu\tau_0)}{\sinh\kappa
y_0}=\lambda_0\beta. \end{equation} The dependence $\tau(y)$ normalized so
that $\tau(0)=0$ reads \be
\tau=\frac{\epsilon(y)}{\kappa\alpha}\ln\left[\frac{\tanh\left(
\kappa(|y|+y_0)/2 \right)}{\tanh (\kappa y_0/2)}\right].\end{equation}

This is valid for any non-zero bulk cosmological constant. The
limit $\Lambda =0$ corresponds to taking $\kappa\to 0$ with
$\kappa\alpha$ fixed. Then \be \tau =
\frac{\epsilon(y)}{\kappa\alpha}\ln\left(\frac{(|y|+y_0)}{y_0}\right),
\end{equation} with \be y_0 = \frac2{|\lambda_0|}\frac{D-2}{D-1}, \end{equation} and \be
Z = \kappa\alpha(|y|+y_0). \end{equation}  Again, the matching condition
gives a negative brane tension.

\setcounter{equation}{0}
\section{Charged RS walls}
The ansatz (\ref{dsw}) of the previous section is actually the
same as for the D8 brane in the massive IIA theory
\cite{BeRo96,BrPe99}, but the form field is different, actually
the D8 brane can be obtained by adding a non-dynamical $A_{[9]}$
potential in ten  dimensions, or, more generally, $A_{[D-1]}$ in
D dimensions. Recently this problem attracted attention in
connection with supersymmetric generalizations of the
Randall-Sundrum setup \cite{AlMeOr00, BeKaPr00,Be01,KaKoLiTs01}.
To clarify the relationship of our F-wall solutions with BPS
brane worlds we construct here the one-brane charged RS type
solution, without a bulk cosmological constant, in any number of
dimensions $D$.

Our action is \ba\label{SD} S_D = \int \left( R   -
\frac12\,\partial_\mu \phi\,\partial^\mu \phi - \frac{1}{2D!}  \,
{\rm e}^{a\phi} \,F_{[D]}^2
\right)\sqrt{-g}\,d^{D}x&&\nonumber\\-\;2\lambda\int  {\rm
e}^{\beta\phi}\sqrt{-g_{D-1}}\,d^{D-1}x   - \;q\int
A_{[D-1]},&&\ea while the ansatz for the space-time metric is
again (\ref{dsw}), so the difference with the previous section is
that the form field now interacts with the brane (the last term
in (\ref{SD})). Clearly, the $D$-form in $D$ dimensions
\be\lb{ansFD} F_{[D]} = f(y)\, dy \wedge dt \wedge dx_1 \wedge
\cdots \wedge dx_{D-2}\end{equation} is trivial and the corresponding term
in the lagrangian acts merely as a dilaton potential. However,
we have to be careful about the jump condition on the source
brane. From the form equation \be\partial_y\left(f(y){\rm
e}^{-(D-1)A+a\phi}\right)=-q\delta(y),\end{equation} it follows that
\be\lb{fwD} f=2b {\rm e}^{(D-1)A - a\phi}, \end{equation} with the
discontinuous parameter \be b=b_+\theta(y)+b_-\theta(-y),\quad
b_\pm={\rm const},\end{equation} and the jump condition \be\lb{bq}
b_+-b_-=-\frac{q}2.\end{equation} Substituting this into the other equations
one arrives at the following system: \ba\lb{ARS} (D-2)[A'' +
(D-1) {A'}^2] &=&-2 b^2{\rm e}^{-a\phi}
-\lambda_0\,\delta(y),\\2(D-1)(D-2){A'}^2 &=& {\phi'}^2 -
4 b^2{\rm e}^{-a\phi},\lb{consR}\\
\phi'' + (D-1)A' \phi' &=& - 2 ab^2{\rm e}^{-a\phi} + 2\beta
\lambda_0\,\delta(y). \lb{phiR} \ea

First we assume the relation 
\be\lb{ba2} \beta=-\frac{a}{2},\end{equation} 
which results from imposing the scale invariance of the action
(\ref{SD}). Again it is
convenient to introduce the function $Z(y)$ via Eq.
(\ref{AZ}). Then, as can be easily seen, the dilaton equation and
Eq. (\ref{ARS}) imply that $\phi$ and $ A$ are proportional
modulo the solution of the homogeneous equation: \be\lb{phiAR}
\phi=a(D-2)A +d(\tau+\tau_0),\quad (d,\;\; \tau_0=\rm const),\end{equation}
where $\tau$ satisfies Eq. (\ref{tz}). The remaining equations are
easily solved in the special case $d=0$. Combining (\ref{phiAR}) with Eq.
(\ref{ARS}) one obtains the following equation for $Z$: \be
\lb{ZRS} Z''+ 2b^2\frac{D-1}{D-2}Z^{-2\nu-1}=
-\lambda_0\frac{D-1}{D-2}\,Z\,\delta(y),\end{equation} with \be
\nu=-1+\frac{a^2(D-2)}{2(D-1)}.\end{equation} The constraint equation
(\ref{consR}) then implies \be\lb{consZ} {Z'}^2-\kappa^2
Z^{-2\nu}=0,\end{equation} where \be \kappa^2=\frac{4 b^2
(D-1)^2}{(D-2)(a^2(D-2)-2D+2)}.\end{equation} For this to be consistent, one
has to impose the condition \be\lb{nupos}
a^2>\frac{2(D-1)}{D-2}\end{equation} ($\nu > 0$), which we assume to hold.
The solution to Eqs. (\ref{ZRS}) and (\ref{consZ}) is readily
obtained, \be Z=\left[\kappa
(1+\nu)(|y|+y_0)\right]^{1/(1+\nu)}.\end{equation} Here the parameter
$\kappa$ is continuous on the brane provided \be\lb{bpm} b_+ = -b_- =
-\frac{q}4, \end{equation} and we have to accommodate for the source term in
Eq. (\ref{ZRS}). This is achieved imposing the following
condition on the parameter $y_0$: \be 4=-\lambda_0 a^2 y_0.\end{equation} In
the case of a single source brane one has to demand positivity
$y_0>0$ to ensure regularity of the metric, so in this case the
appropriate brane tension is negative. However, in a many-brane
setup such as in \cite{KaKoLiTs01}  one can achieve a non-singular metric
including positive tension branes as well.

A more general solution can be derived by relaxing the condition
(\ref{ba2}). Putting
\be\lb{bpsi} B \equiv A', \quad \psi \equiv \frac{\kappa}{D-1}{\rm
e}^{-a\phi/2}, \end{equation} and combining Eqs. (\ref{ARS}) and (\ref{consR})
we obtain, outside the brane, \be\lb{phip} \phi' =
a(D-2)\psi\frac{dB}{d\psi}.  \end{equation} Inserting this in Eq.
(\ref{consR}) yields the Lagrange-type differential equation \be
(1+\nu)\psi^2\left(\frac{dB}{d\psi}\right)^2- B^2 - \nu\psi^2 = 0.
\end{equation} To solve this, put \be B = t\psi, \quad \frac{dB}{d\psi} = \pm
f(t) \equiv \pm\frac{q}{p}\sqrt{t^2 + p^2} \end{equation} with
\be p^2 = \nu, \quad q^2 = \frac{\nu}{1+\nu} \end{equation}
($p > 0$, $q > 0$). The case $d = 0$ corresponds to $t
\equiv \pm 1$, $f \equiv 1$. In the present case, \be
\frac{d\psi}{\psi} = \frac{dt}{\pm f(t)-t}, \end{equation} which is
integrated, for $q \neq p$, by \be
\ln\left(\frac{\psi_{\pm}}{\psi_0}\right) =
\frac{p}{q^2-p^2}\left(p\ln|q\mp p\sin\theta|
\pm q\ln(1+\sin\theta) -
(p\pm q)\ln(\cos\theta)\right), \end{equation} where we have put $t
= p\tan\theta$ ($-\pi/2 < \theta < \pi/2$). Also, it follows
from the definitions (\ref{bpsi}) and from Eq. (\ref{phip}) that
\be\lb{Aythe} dA = -\frac2{(D-2)a^2}\left(\frac{d\psi}{\psi} \mp
\frac{p}{q}\frac{d\theta}{\cos\theta}\right), \quad dy =
-\frac{2p
d\theta}{(D-2)a^2q(q\mp p\sin\theta)\psi}. \end{equation}

The resulting spacetime metric after suitable rescaling of $ x^a$
is
\begin{multline}
ds_{\pm}^2 = \left(\frac{(1+\sin\theta)^{\pm\gamma}
|1\mp\gamma\sin\theta|}{(\cos\theta)^{1\pm\gamma}}\right)
^{\frac{2/(D-1)}{\gamma^2-1}}\eta_{ab}\,dx^a\,dx^b +
\\+
\frac1{((D-1)p
\psi_0)^2}\left(\frac{(1+\sin\theta)^{\pm\gamma}|1\mp\gamma\sin\theta|}
{(\cos\theta)^{\pm\gamma(1\pm\gamma)}}\right)^{\frac2{\gamma^2-1}}\,d\theta^2,
\end{multline}
with $\gamma = p/q = \sqrt{1+\nu}$, while the dilaton
function and form potential read 
\ba 
{\rm e}^{-a\phi/2} & = & \frac{D-1}{\kappa}
\psi_0\bigg[\frac{q(1+\sin\theta)^{\pm1}}{(\cos\theta)^{\gamma\pm1}}
(1\mp\gamma\sin\theta)^{\gamma}\bigg]^{-\frac{\gamma}{\gamma^2-1}}, \\
A_{[D-1]} & = &
\pm\frac{\psi_0}{b(D-2)}\bigg(\frac{\gamma}{\sqrt{\gamma^2-1}}\bigg)
^{\frac{2\gamma}{\gamma^2-1}-1}\,(1\mp\gamma\sin\theta)^{-1}.
\ea
Note that $ds_{\pm}^2(-\theta) = ds_{\mp}^2(\theta)$ (the same
being true for the dilaton $\phi$).

For $\gamma < 1$ ($\nu < 0$), this metric is singular for $\theta
= \pi/2$ and for $\theta = -\pi/2$. Putting $\theta =
\epsilon(\pi/2-u)$, we find that for $u \to 0$, $y_{\pm} \sim
u^{\frac1{1\mp\epsilon\gamma}}$ and the metric behaves as \be
ds^2 \sim |y|^{2/(D-1)}\eta_{ab}\,dx^a\,dx^b + dy^2  \quad (y \to
0). \end{equation} Because there are two singularities, it is not possible
to obtain a regular solution with a single source brane. On the
other hand, for $\gamma > 1$ ($\nu > 0$), i.e. when the condition
(\ref{nupos}) holds, the singularity at $\theta = \pm\pi/2$ is
sent off to infinity, while a new singularity appears at $\theta
= \pm\theta_0$ with $\sin\theta_0 = 1/\gamma$. So the regular RS
type solution is given by \ba (ds^2, \phi, f) & =
(ds_+^2, \phi_+, f_+)(\theta) \quad & (\theta_1 < \theta < \pi/2) \\
& = (ds_-^2, \phi_-, f_-)(\theta) \quad & (-\pi/2 < \theta <
-\theta_1) \ea with $\theta_0 < \theta_1 < \pi/2$. The continuity
of the metric and of the dilaton on the brane $y = 0$ ($\theta =
\pm\theta_1$) is ensured by the above-mentioned symmetry together
with the relation (\ref{bpm}) which fixes the strength of the
form field, while the full equation (\ref{ARS}) gives for the
jump of the connexion \be 2(D-2)A_+'(\theta_1) =
2(D-2)p\tan\theta_1\psi_+(\theta_1) = -\lambda_0, \end{equation} so
that the brane tension is again negative. Finally, the comparison
of Eqs. (\ref{ARS}) and (\ref{phiR}) yields \be \beta =
-\frac{a}2\frac{f(\theta_1)}{t(\theta_1)} = -
\frac{a}2\frac{\sin\theta_0}{\sin\theta_1} > - \frac{a}2 \end{equation} (it
follows that for $\beta = -a/2$ the $d = 0$ solution is the unique
regular solution).

\section{Conclusion }
In this paper we have presented two classes of solutions to gravity
coupled to a dilaton and to an antisymmetric form in arbitrary
dimensions. These two classes lie at the limits of the sequence 
of fluxbranes  of different dimensionalities

The first class is that of zero fluxbranes supported by
a form of rank $D-2$. Presumably (the Euclidean continuations
of) such branes  are entitled to interact with instantons. 
We have shown that such 
solutions may exist with the transverse spaces $S_k\times R^{(D-k-2)}$,
with $1\leq k \leq D-2$, and in the case $k=1$ 
we obtained a complete analytic solution
containing four physical  parameters. One of these parameters is
connected with the form field strength, while the other parameters 
are associated with pointlike singularities on the
regular Melvin-type background. Solutions of an entirely new type are
obtained in the case of a spherically symmetric transverse space
$k=D-2$. The full metric then consists of an infinite number of disjoint
sectors, each of which is shown to be geodesically complete.
For other values of $k$ regular solutions exist which can be found 
numerically.
 
At the other extreme of the supergravity fluxbrane sequence lies the
F-wall  of codimension one. Actually, this is a domain wall
supported by the scalar field. Such a solution is easy to find in a
closed form which reveals its singularity on a hyperplane of dimension
$(D-1)$. We explored whether one can regularize the solution by
orbifolding the space at some point away from the singularity. This
leads to some new brane world setup with two scalar fields. In the case
where only one brane is present, the space is non-singular if the
brane tension is negative (in a multibrane setup positive tension branes 
will be present too). The general solution 
was found with a bulk cosmological
constant added. It is worth noting that in this setup the brane 
does not carry a Ramond-Ramond charge.

We have also constructed new brane-world solutions with charged branes,
assuming an arbitrary coupling of the dilaton to the brane. 
The usual supersymmetric
brane-world corresponds to a particular value of this constant. In the
general case the solution turns out to be strikingly different. We found 
conditions for its regularity, and, as a by product, have shown uniqueness
of the supersymmetric brane-world.

\section*{Acknowledgements}
One of the authors  thanks LAPTH (Annecy-le-Vieux) for hospitality and support.
He also acknowledges support from the RFBR under grant 00-02-16306.

\end{document}